# The Study on the Effects of Chromaticity and Magnetic Field Tracking Errors at CSNS/RCS[*]

XU Shou-Yan(许守彦)[&], WANG Sheng(王生), AN Yu-Wen(安宇文), LI Zhi-Ping(李志平)

Institute of High Energy Physics (IHEP), Beijing, 100049, China

**Abstract:** The Rapid Cycling Synchrotron (RCS) is a key component of the China Spallation Neutron Source (CSNS). For this type of high intensity proton synchrotron, the chromaticity, space charge effects and magnetic field tracking errors between the quadrupoles and the dipoles can induce beta function distortion and tune shift, and induce resonances. In this paper the combined effects of chromaticity, magnetic field tracking errors and space charge on beam dynamics at CSNS/RCS are studied systemically. 3-D simulations with different magnetic field tracking errors are performed by using the code ORBIT, and the simulation results are compared with the case without tracking errors.

## 1 Introduction

The China Spallation Neutron Source (CSNS) is an accelerator-based facility. It operates at 25 Hz repetition rate with the design beam power of 100 kW. CSNS consists of a 1.6-GeV Rapid Cycling Synchrotron (RCS) and an 80-MeV linac. RCS accumulates 80 MeV injected beam, and accelerates the beam to the design energy of 1.6 GeV, and extracts the high energy beam to the target. The lattice of the CSNS/RCS is triplet based four-fold structure. Table 1 shows the main parameters for the lattice [1] [2].

The preferred working point of CSNS/RCS is (4.86, 4.78) which can avoid the major low-order structure resonances. But because of the chromatic tune shift [3], space-charge incoherent tune shift and the tune shift caused by magnetic field tracking errors between the quadrupoles and the dipoles, some structure resonances are unavoidable. The chromaticity, space charge effects and magnetic field tracking errors can also induce beta function distortion, and influence the transverse acceptance and the collimation efficiency of the collimation system. In such a situation, a clear understanding of the effects of magnetic field tracking errors, space charge, and the chromaticity on beam dynamics at CSNS/RCS is an important issue.

In this paper the effects of chromaticity, magnetic field tracking errors and space charge on beam dynamics at CSNS/RCS are studied systemically. 3-D simulations are performed introducing magnetic field tracking errors and space charge effects. The combined effects of chromaticity, magnetic field tracking errors and space charge on the beam dynamics for CSNS/RCS are discussed.

Table 1: Main parameters of the CSNS/RCS Lattice.

| Circumference (m) | 227.92 |
| --- | --- |
| Super period | 4 |
| Betatron tunes (h/v) | 4.86/4.78 |
| RF harmonics | 2 |
| Injection energy (MeV) | 80 |
| Accumulated particles per pulse | $1.56\times 10^{13}$ |
| Transverse acceptance (μπm.rad) | >540 |
| Acceptance of the primary Collimators (H/V, μπm.rad) | 350 |
| Acceptance of the secondary Collimators (H/V, μπm.rad) | 400 |

## 2 Effects on Lattice function

The natural chromaticity of the CSNS/RCS lattice is (-4.3, -8.2), which can produce the tune shift of (±0.04, ±0.08) for the momentum spread of $\Delta p/p=\pm 0.01$ (The maximum momentum spread during acceleration according to the simulation results). The dependence of the Beta functions on the momentum spread along a super-period without chromatic correction is shown in Fig. 1.

In the case of uniform distribution in transverse

________________________
*: supported by National Natural Sciences Foundation of China
(No. Y2113A005C)
&: xusy@ihep.ac.cn

direction, the incoherent tune shift due to space charge effects can be expressed as:

$$\Delta \upsilon = -\frac{r_p N}{8\pi \varepsilon_{rms} \beta^2 \gamma^3 B_f} \quad (1)$$

where $r_p = 1.53 \times 10^{-18}$ m is the classical proton radius, N is the accumulated particles, $\varepsilon_{rms}$ is the un-normalized RMS. emittance, $B_f$ is the longitudinal bunching factor, $\beta$ and $\gamma$ are the relativistic Lorentz factors. For CSNS/RCS, the longitudinal bunching factor, just after the injection painting, is about 0.32. With the energy of 80MeV, the space charge induced incoherent tune shift is about 0.26 for the case $\varepsilon_{rms} = 60\pi\mu$m.rad [4]. For the actual beam, which deviates from the uniform distribution, the incoherent tune shift for the particles in the beam core may be much greater than 0.26.

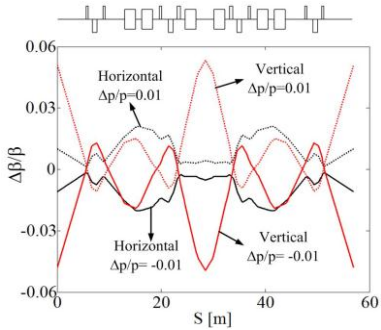

Fig. 1 The dependence of the Beta functions on the momentum spread along a super-period without chromatic correction ($\Delta\beta = \beta_{\Delta p \neq 0} - \beta_{\Delta p = 0}$).

For the rapid cycling synchrotron, there may be magnetic field tracking errors between the quadrupoles and the dipoles because of saturations of the fields. The field tracking errors can induce beta function distortion and tune shift. The tune shift induced by magnetic field tracking errors can be expressed as:

$$\Delta \upsilon = \frac{1}{4\pi} \oint_C \Delta K \cdot \beta(s) \cdot ds = -\frac{\Delta K}{K} \xi_{natural} \quad (2)$$

Where $\Delta K$ is the effective focusing error induced by mismatch of dipole field and quadrupole field, $\xi_{natural}$ is the natural chromaticity. The CSNS/RCS focusing structure consists of 48 quadrupole magnets, which can be divided into four types named QA, QB, QC and QD in this paper. In order to assure a close tracking between the dipole and the quadrupole, the magnets for CSNS/RCS are designed so that the saturations of the fields are within 1.5% for dipoles, within 2% for QA and within 1.5% for the other three types of quadrupole magnets [5]. To reduce the magnetic tracking errors, higher frequency waves can be used to compensate the field saturation for both dipoles and quadrupoles.

2% tracking errors for QA and 1.5% tracking errors for QB, QC, QD are considered first preparing for the worst case. The tune shift induced by magnetic field tracking errors is (-0.087, -0.095). The effects of magnetic field tracking errors on Beta functions are shown in Fig. 2. Because of the distortion of Beta functions, the physical acceptance and the acceptances of the collimators are changed with the collimators aperture unchanged.

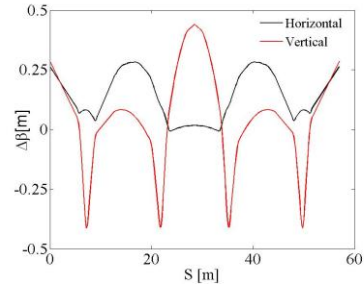

Fig. 2. The effects of magnetic field tracking errors on Beta functions ($\Delta\beta = \beta_{with-error} - \beta_{no-error}$).

In the measurements of the prototype quadrupole magnet of CSNS/RCS, it is confirmed that the tracking errors can be adjusted within 0.1% by compensating by using higher frequency waves [6]. In the actual operations, all the 48 quadrupole magnets are divided into five groups. Every group includes 8 or 16 quadrupole magnets distributed in different spuper period. The quadrupole magnets in the same group are powered by the same power supply and have the same type and magnetic field. Considering the difference between the quadrupole magnets in the same group, which is designed to be less than 0.2%, the magnetic tracking errors can be up to 0.3% after compensating using high frequency waves. The combined effects of chromaticity and magnetic field tracking errors on tunes are shown in Fig. 3. In Fig. 3, different tunes are obtained with different Δp/p and 20 groups of magnetic field tracking errors, which vary in the range of (-0.3%, 0.3%). Considering the space charge tune shift, which is larger than 0.26, the tunes can be close to the resonances $2\nu_x-2\nu_y=0$ and $2\nu_y=9$, which are dangerous resonances for CSNS/RCS [4] [7].

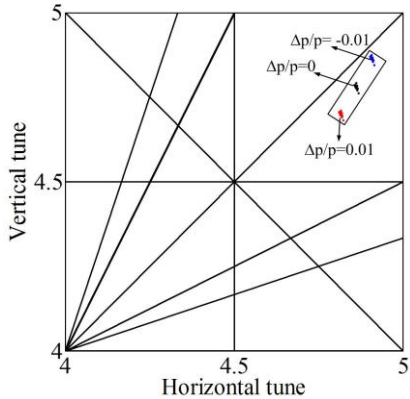

Fig. 3 The combined effects of chromaticity and magnetic field tracking errors on tunes.

## 3  Simulation results

3-D simulations with different magnetic field tracking errors are done by using the code ORBIT, and the simulation results are compared with the case without tracking errors. Two types of magnetic field tracking errors are simulated:

Case-1: random errors in the range of (-0.3%, 0.3%) for all 48 different quadrupole magnets;

Case-2: random errors in the range of (-0.3%, 0.3%) for different quadrupole groups, which means that the quadrupole magnets in the same group have the same errors.

Without space charge effects, there is no apparent emittance growth and beam loss with and without magnetic field tracking errors.

The simulation results for case-1 with space charge effects are shown in Fig. 4 and are compared with the case without tracking errors. Without magnetic field tracking errors, the beam loss is less than 1%, and the uncontrolled beam loss is less than 0.1% in the early 5ms. With the random errors in the range of of (-0.3%, 0.3%) for all 48 different quadrupole magnets, there is serious beam loss during acceleration, which is up to 6.5% in the early 5ms. The uncontrolled beam loss is about 0.5%. Different magnetic field tracking errors are simulated and compared. The dependence of beam loss on the magnetic field tracking errors is shown in Fig. 5. The beam loss gradually increases with the increase of magnetic field tracking errors.

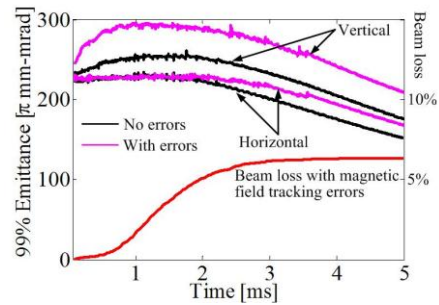

Fig. 4 The simulation results for case-1 with space charge effects.

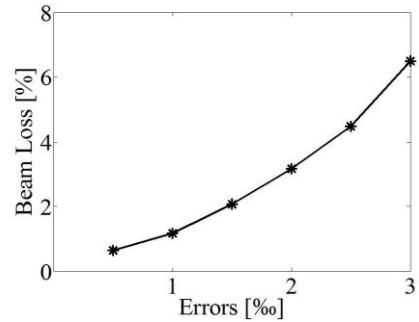

Fig. 5 The beam loss in the early 5ms with different magnetic field tracking errors.

The simulation results for case-2 with space charge effects show that there is no serious beam loss, which is about 1.3% in the early 5ms. The time evolutions of the 99% eimmtances are shown in Fig. 6.

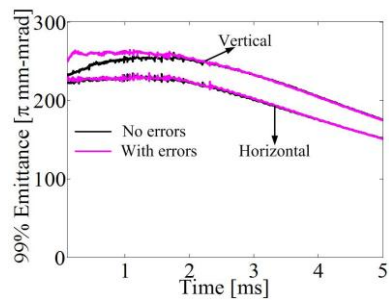

Fig. 6 The simulation results for case-2 with space charge effects.

Compared the simulation results for Case-1 and Case-2, there is much more beam loss and emittance growth for Case-1. The explanation might be that the super period structure is distorted for Case-1, and the generic resonances change to structural resonances, which are stronger than generic ones.

For CSNS/RCS, the magnetic field tracking errors are within 0.1% after compensating by using higher frequency waves, and the difference between the quadrupole magnets in the same group is designed to be less than 0.2%. So, the case with the random errors in the range of (-0.1%, 0.1%) for different quadrupole groups and the errors in the range of (-0.2%, 0.2%) for different quadrupole magnets in the same group is more close to the actual case. The simulation results for this case are shown in Fig. 7, which shows the dependence of beam loss on the stored particles. $N_0$ is the designed number of stored particles, which corresponds to the beam power of 100 kW. There is apparent impact of stored particles on the beam loss. The beam loss increases rapidly with the increase of the number of stored particles when the number of stored particles is larger than $0.85N_0$.

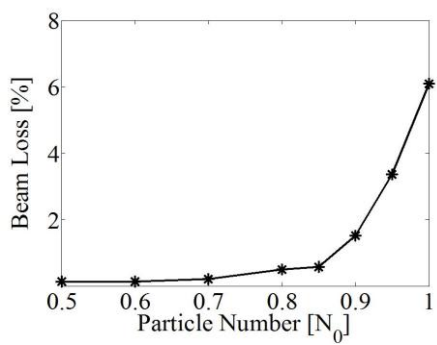

Fig. 7 The beam loss with different stored particles in the early 5ms with the random errors in the range of (-0.1%, 0.1%) for different quadrupole groups and the random errors in the range of (-0.2%, 0.2%) for different quadrupole magnets in the same group.

## 4  Summary

In this paper the combined effects of chromaticity, magnetic field tracking errors and space charge on beam dynamics at CSNS/RCS are studied systemically. The space charge induced incoherent tune shift is about -0.26, and the tune spread produced by the chromaticity is about (±0.04, ±0.08). Without compensating by using higher frequency waves, the tune shift induced by magnetic field tracking errors can get up to (-0.087, -0.095). The largest tune shift can get up to -0.4 on the combined effects of chromaticity, magnetic field tracking errors and space charge. To reduce the magnetic field tracking errors, higher frequency waves are used to compensate the field saturation. Considering the difference between the quadrupole magnets in the same group, the tracking errors are less than 0.3% after compensating by using higher frequency waves. The simulation results show that the magnetic field tracking errors have an apparent effect on beam loss, and the random errors for different quadrupole magnets in the same group are more dangerous than the errors for different quadrupole groups which does not distort the super period structure. So the difference between the quadrupole magnets in the same group should be controlled as small as possible.